\documentclass{emulateapj}
\usepackage{apjfonts}
\usepackage{epsfig}

\newcommand\Dk{\Delta_{\kappa}^2 (\ell)}

\begin{document}

\title{Impact of Dark Matter Substructure on the Matter and Weak
Lensing Power Spectra}

\author{ Bradley Hagan\altaffilmark{1}, Chung-Pei Ma\altaffilmark{2},
Andrey V. Kravtsov\altaffilmark{3} }

\altaffiltext{1}{Department of Physics, University of California,
  Berkeley, CA 94720} 
\altaffiltext{2}{Department of Astronomy, University of California,
  Berkeley, CA 94720} 
\altaffiltext{3}{Dept. of Astronomy and
  Astrophysics, Enrico Fermi Institute, Kavli Institute for
  Cosmological Physics, The University of Chicago, Chicago, IL~60637}

\begin{abstract}

We explore the effect of substructure in dark matter halos on the
power spectrum and bispectrum of matter fluctuations and weak lensing
shear.  By experimenting with substructure in a cosmological
$N=512^3$ simulation, we find that when a larger fraction of the host
halo mass is in subhalos, the resulting power spectrum has less power
at $1 \la k \la 100 h$ Mpc$^{-1}$ and more power at $k \ga 100 h$
Mpc$^{-1}$.  We explain this effect using an analytic halo model
including subhalos, which shows that the $1 \la k \la 100 h$
Mpc$^{-1}$ regime depends sensitively on the radial distribution of
subhalo centers while the interior structure of subhalos is important
at $k \ga 100 h$ Mpc$^{-1}$.  The corresponding effect due to
substructures on the weak lensing power spectrum is up to $\sim
11\%$ at angular scale $l \la 10^4$.  Predicting the nonlinear power
spectrum to a few percent accuracy for future surveys would therefore
require large cosmological simulations that also have exquisite
numerical resolution to model accurately the survivals of dark matter
subhalos in the tidal fields of their hosts.

\end{abstract}


\section{Introduction}
\label{sec:intro}

One phenomenon to emerge from $N$-body simulations of increasingly
higher resolution is the existence of substructure (or subhalos) in
dark matter halos (e.g., Tormen et al. 1998; Klypin et al. 1999b;
Moore et al. 1999; Ghigna et al. 2000).  These small subhalos, relics
of hierarchical structure formation, have accreted onto larger host
halos and survived tidal forces.  Depending on their mass, density
structure, orbit, and accretion time, the subhalos with high central
densities can avoid complete tidal destruction although many lose a
large fraction of their initial mass.  These small and dense dark
matter substructures, however, are prone to numerical artifacts and
can be disrupted due to insufficient force and mass resolution.
Disentangling these numerical effects from the actual subhalo dynamics
is an essential step towards understanding the composition and
formation of structure.  Quantifying the effects due to dark matter
substructure is also important for interpreting weak lensing surveys,
which are sensitive to the clustering statistics of the overall
density field.  The level of precision for which surveys such as SNAP
are striving (Massey et al. 2004) suggests that theoretical
predictions for the weak lensing convergence power spectrum need to be
accurate to within a few percent over a wide of range of scales
(e.g. Huterer \& Takada 2005).  At this level, subhalos may contribute
significantly to the nonlinear power spectrum because they typically
constitute about 10\% of the host mass.

In the sections to follow, we examine the effects of substructure on
the matter and weak lensing power spectra with two methods.  In \S~2
we use the result of a high resolution $N$-body simulation and
quantify the changes in the power spectra when we smooth out
increasing amounts of substructures.  Our other approach, detailed in
\S~3, is to incorporate substructure into the analytic halo model.
The results are dependent on the parameters used in the model, but
they provide useful physical insight into the results from $N$-body
simulations.  We summarize and discuss the results in \S~4.

\section{Substructure in Simulations}
 
We use the outputs of a cosmological dark-matter-only simulation that
contains a significant amount of substructure.  This simulation is a
concordance, flat $\Lambda$CDM model: $\Omega_m = 1-\Omega_\Lambda =
0.3,\ h = 0.7$ and $\sigma_8 = 0.9$.  The box size is $120 h^{-1}$
Mpc, the number of particles is $512^3$, and the particle mass is
$1.07\times 10^9 h^{-1} M_\odot$.  The simulation uses the Adaptive
Refinement Tree $N$-body code (ART; Kravtsov et al. 1997; Kravtsov
1999) to achieve high force resolution in dense regions.
In this particular run the volume is initially resolved with a
$1024^3$ grid, and the smallest grid cell found at the end of the
simulation is $1.8 h^{-1}$ kpc.  The actual resolution is about twice
this value (Kravtsov et al. 1997).  More details about the simulation
can be found in Tasitsiomi et al. (2004).

To quantify the effects of subhalos on the matter and weak lensing
power spectra, we first identify the simulation particles that
comprise subhalos within each halo.  This is achieved using a version
of the Bound Density Maxima algorithm (Klypin et al. 1999a), which
identifies all local density peaks and therefore finds both halos and
subhalos.  It identifies the particles that make up each of the peaks
and removes those not bound to the corresponding halo.  As a
controlled experiment, we then smooth out the subhalos within the
virial radius of each host halo by redistributing these subhalo
particles back in the smooth component of the host halo according to a
spherically-symmetric NFW profile (Navarro, Frenk, \& White 1996).
For the concentration parameter $c$ of the profile, we do not use the
fitting formulae (e.g., Bullock et al. 2001; Dolag et al. 2004) but
instead fit each host halo individually to take into account the
significant halo-to-halo scatter in $c$.  We therefore smooth over the
subhalos and increase the normalization but not the shape of the
spherically averaged profile of the smooth component to accommodate
the mass from the subhalo component.

This smoothing procedure also serves as a simple model for the effects
of resolution on the abundance of subhalos in simulations, in which
the lack of sufficient resolution will cause an incoming small halo to
be disrupted quickly and lose most of its particles over its
short-lived orbit.  We quantify this effect by experimenting with
different cut-offs on the subhalo mass:  subhalos with masses below
the cut-off are removed and have their particles spread over the host
halo; subhalos with masses above the cut-off are left alone.
Increasing mass cut-offs should roughly mimic increasingly lower
resolution simulations because only higher mass subhalos will stay
intact in the halo environment.

We then calculate the matter fluctuation power spectrum $P(k)$, the
Fourier transform of the 2-point correlation function, and the matter
bispectrum $B(k_1,k_2,k_3)$, the Fourier transform of the 3-point
correlation function.  In order to compute $P$ and $B$ at large
$k$ without using an enormous amount of memory, we subdivide the
simulation cube into smaller cubes and stack these on top of each
other (called "chaining the power" in Smith et al.~2003).  A typical
stacking level used is 8, meaning that we subdivide the box into $8^3$
cubes and stack these.  We use stacked spectra for the high-$k$ regime
and unstacked spectra for low-$k$.  Finally, we subtract shot noise
($\propto 1/N$) from the outputted spectra to eliminate discreteness
effects.

\begin{figure}[tp]
\centerline{
\epsfig{file=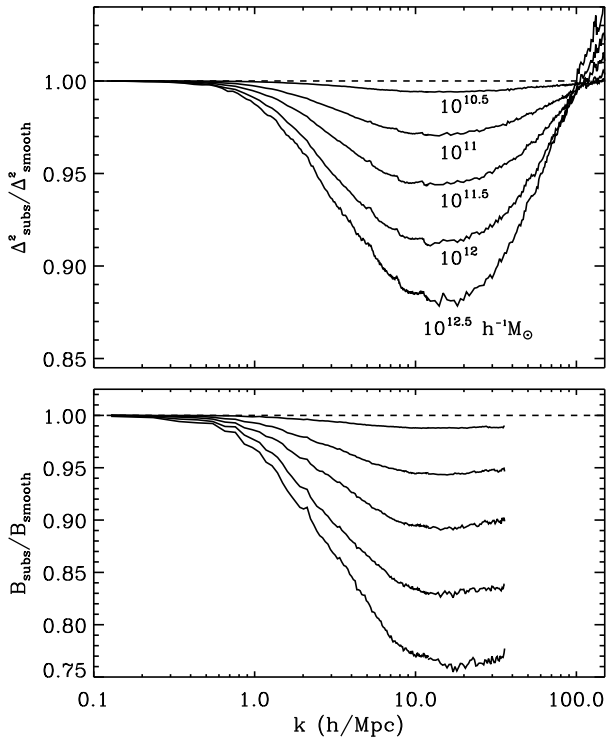, height=4.5in}
}
\caption{Effects of dark matter substructure on the matter fluctuation
power spectrum (top) and the equilateral bispectrum (bottom) of a
$N=512^3$ cosmological simulation.  Plotted is the ratio of the
original spectrum to that with a subset of the substructures smoothed
out.  The curves (from top down) correspond to increasing subhalo mass
cut-offs, below which the mass in the subhalos is redistributed
smoothly back into the host halo.  The bottom panel is plotted to a
lower $k$ because $B(k)$ becomes too noisy.  The effects due to
substructures are up to $\sim 12$\% in $\Delta^2$ and 24\% in $B$.}
\label{fig:pk}
\end{figure}

Fig.~\ref{fig:pk} shows the effects of substructure on the {\em
dimensionless} power spectrum $\Delta^2(k) \equiv k^3 P(k) / (2\pi^2)$
and the equilateral bispectrum $B(k_1)$ ($k_1=k_2=k_3$).  Plotted is
the ratio of the spectrum from the raw ART output divided by the
spectrum from the altered data.  The altered data have no subhalos with
masses below the labeled mass cutoff.  The mass in the removed
subhalos has been redistributed smoothly into the host halo as
described above.  The deviation from the original spectrum becomes
larger as the cutoff is increased because more subhalos have been
smoothed out.  For a given cutoff, the figure shows that a simulation
with dark matter substructures (such as the raw ART output) has ${\it
more}$ power at $k>100$ {\em h} Mpc$^{-1}$ and {\it less} power at $1
\lesssim k \lesssim 100\ h$ Mpc$^{-1}$ than a simulation with smoother
halos.  We believe these opposite behaviors reflect the two competing
factors present in our numerical experiments: {\it removal} of mass
within subhalos, which affects scales comparable to or below subhalo
radii (and hence $k \ga 100 h$ Mpc$^{-1}$), and {\it addition} of this
mass back into the smooth component of the halo, which affects the
larger scales of $1 \lesssim k \lesssim 100\,h$ Mpc$^{-1}$.  The ratio
approaches unity for $k \la 1\,h$ Mpc$^{-1}$ simply because the mass
distribution on scales above individual host halos is unaltered.  We
will examine these effects further in the context of the halo model in
\S~3.

For a given curve in Fig.~\ref{fig:pk}, we have also calculated the
contributions from subhalos in host halos of varying masses to
quantify the relative importance of cluster versus galactic host
halos.  For the $10^{12.5} h^{-1} M_\odot$ curve, e.g., we find that
smoothing over the subhalos in host halos above $10^{14}$ and $10^{13}
M_\odot$ account for 5\% and 10\% in the total 12\% dip seen in
Fig.~\ref{fig:pk}, respectively.  For the $10^{11.5} h^{-1} M_\odot$
cutoff, the numbers are 2\% and 5\% of the total 6\% dip.

The halos found in $N$-body simulations are generally triaxial.  When
we redistribute the subhalo particles, however, we assume for
simplicity a spherical distribution.  This assumption makes the
altered halos slightly rounder.  One can estimate how this effect
changes the power spectrum by using the halo model without
substructure.  Smith \& Watts (2005) incorporated a distribution of
halo shapes found by Jing \& Suto (2002) from cosmological simulations
into the halo model (ignoring the substructure contribution).
Compared with the case where all the halos are spherical, they
observed a peak decrement in the power spectrum of about $4\%$ for $k
\approx 1$ Mpc$^{-1}$.  The corresponding effect in our calculations
would be much smaller since we redistribute only the subset of
particles that belong to subhalos into the rounder shape (e.g., about
10\% of all particles in the case where the cutoff was $10^{12.5}
h^{-1} M_\odot$).  Thus, by extending the results of Smith \& Watts,
we expect the spurious rounder halos in our study to account for less
than $0.5 \%$ of the total $12\%$ drop.

\begin{figure}[t]
\centerline{
\epsfig{file=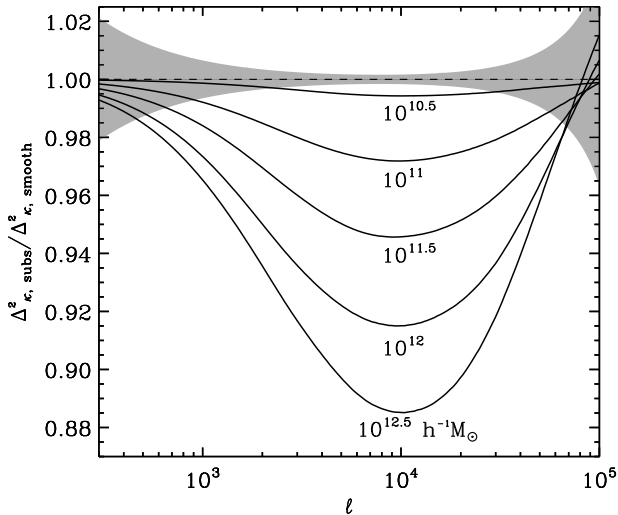, height=3.0in}
}
\caption{Effects of substructure on the weak lensing convergence power
spectrum for the same subhalo mass cutoffs as in Fig.~1.  The sources
are assumed to be at redshift 1.  The gray band is the 1-$\sigma$
statistical error assuming Gaussian fields.  We take $f_{sky}=0.25,\ \gamma_{rms} =
0.2,\ \bar{n}=100/$arcmin$^2$, and a band of width $\ell/10$. }
\label{fig:weak}
\end{figure}

Fig.~\ref{fig:weak} shows the weak lensing convergence power spectrum
$\Delta^2_\kappa(l)$ corresponding to the matter power spectrum
$\Delta^2(k)$ in Fig.~\ref{fig:pk}.  It is calculated from
$\Delta^2(k)$ using Limber's approximation and assumption of a flat
universe:
\begin{equation}
     \Dk = \frac{9\pi}{4 \ell} \Omega_m^2 \left(\frac{H_0}{c}\right)^4
        \int^{\chi_{max}}_0 \chi^3 d\chi
        \frac{W^2(\chi)}{a^2(\chi)}\Delta^2(k=\ell/\chi,a) \,.
\end{equation}
Here $\chi$ is the comoving radial distance, and the weak lensing
weight is $W(\chi) = \int^{\chi_{max}}_{\chi} d\chi' p(\chi')
(\chi'-\chi)/\chi'$, where $p(\chi)$ is the distribution of source
galaxies such that $\int p(\chi)d\chi = 1$ (see, e.g., Bartelmann \&
Schneider 2001).  Here we assume for simplicity that all sources are
at one redshift ($z=1$) and use the $z=0$ simulation output to
estimate $\Delta^2(k)$.  Host halos at higher redshift may have a
larger fraction of their mass in substructures because the typical
subhalo accretion epoch would be more recent and there would be less
time for tidal disruption. The effects of $z\sim 0.5$ substructure may
therefore be somewhat larger than shown here for $z=0$, although we do
not expect the change to be significant.

The gray band in Fig.~\ref{fig:weak} marks the extent of uncertainties
from sample variance and shape noise in weak lensing measurements
assuming {\em Gaussian} density fields (Kaiser 1998):
\begin{equation}
\label{eqn:noise}
   {\sigma(\Delta_\kappa^2) \over \Delta_\kappa^2} = 
   \sqrt{2 \over (2\ell +1) f_{sky}} \left(1 + {\ell^2 \gamma^2_{rms} 
    \over 2 \pi \bar{n} \Delta_\kappa^2} \right) \,,
\end{equation}
where $f_{sky}$ is the fraction of the sky surveyed, $\gamma_{rms}$ is
the rms ellipticity of galaxies, and $\bar{n}$ is the number density
of galaxies on the sky.  The error is dominated by the sample variance
on large scales (first term in eq.~[\ref{eqn:noise}]) and the "shape
noise" on small scales.  Our assumption of Gaussianity is not
applicable for the angular scales shown in the plot because the scales
plotted are near or below the size of individual halos, but the errors
shown should be a useful reference and have been used in previous
studies.  A reliable estimate of the error would presumably require a
ray tracing calculation which is beyond the scope of this paper.

\section{Substructure in the Halo Model}

To gain a deeper understanding of the simulation results in
Figs.~\ref{fig:pk} and \ref{fig:weak}, we use the semi-analytic halo
model to build up the nonlinear power spectrum from different kinds of
pairs of mass elements that may occur in halos (e.g. Ma \& Fry 2000;
Peacock \& Smith 2000; Seljak 2000; Scoccimarro et al 2001).  The
original halo model assumes that all mass resides in virialized,
spherical halos without substructures.  One can then build the matter
power spectrum from the different kinds of pairs of particles that
contribute to the 2-point clustering statistics by writing $P(k) =
P_{1h}(k) + P_{2h}(k)$, where the 1-halo term $P_{1h}$ contains
contributions from particle pairs where both particles reside in the
same halo, and the 2-halo term $P_{2h}$ is from pairs where the two
particles reside in different halos.  The 1-halo term is a
mass-weighted average of single halo profiles and dominates on the
scales of interest ($k \ga 1 h$ Mpc$^{-1}$) in Fig.~\ref{fig:pk}
because close pairs of particles are more likely to be found in the
same halo.  The 2-halo term is closely related to the linear power
spectrum and is important only at large separation (i.e. small $k$)
where a pair of particles is more likely to be found in two distinct
halos.  Similarly, the bispectrum can be constructed from the
different classes of triplets of particles (see, e.g., Ma \& Fry
2000).

The original halo model can be readily extended to take into account a
clumpy subhalo component in an otherwise smooth host halo.  Sheth \&
Jain (2003), e.g., decompose the original 1-halo term into $P_{1h} =
P_{ss} + P_{sc} + P_{1c} + P_{2c}$, where "s" denotes smooth and "c"
denotes clump.  The smooth-smooth term, $P_{ss}$, arises from pairs of
particles that both belong to the smooth component of the same host
halo.  This term is identical to the original 1-halo term except for
an overall decrease in amplitude by the factor $(1-f)^2$, where $f$ is
the fraction of the total halo mass that resides in subhalos.  The
smooth-clump term, $P_{sc}$, is due to having one particle in a
subhalo (clump) and the other in the host halo (smooth).  The 1- and
2-clump terms, $P_{1c}$ and $P_{2c}$, come from having both particles
in the same subhalo and in two different subhalos, respectively.
Explicitly,
\begin{eqnarray} 
P_{ss}(k) &=& {(1-f)^2 \over {\bar\rho}^2} \int dM\ N(M) M^2 U^2(k,M) \\
P_{sc}(k) &=& 2{(1-f) \over {\bar\rho}^{2}} \int dM\ N(M) M\, 
     U(k,M)U_c(k,M)\nonumber \\
     & \times & \int dm\ n(m,M) \, m \, u(k,m)  \label{haloterms}\\
P_{1c}(k) &=& {1 \over{\bar\rho}^{2}} \int dM N(M) \int dm\ n(m,M) m^2\, 
     u^2(k,m) \\
P_{2c}(k)  & = & {1 \over{\bar\rho}^{2}}\int dM N(M)  U_c^2(k,M) \nonumber \\
& \times & \left[ \int dm n(m,M) m u(k,m) \right]^2
\end{eqnarray} 
where $U(k,M)$, $u(k,m)$, and $U_c(k,M)$ are the Fourier transforms of
the host halo radial density profile, the subhalo radial density
profile, and the radial distribution of subhalo centers, respectively.
$N(M) dM$ gives the number density of host halos with mass $M$, and
$n(m,M)dm$ gives the number density of subhalos of mass $m$ inside a
host halo of mass $M$.  A similar expression can be written down for
the 2-halo term $P_{2h}$, which is also included in our calculations.

\begin{figure}[t]
\centerline{
\epsfig{file=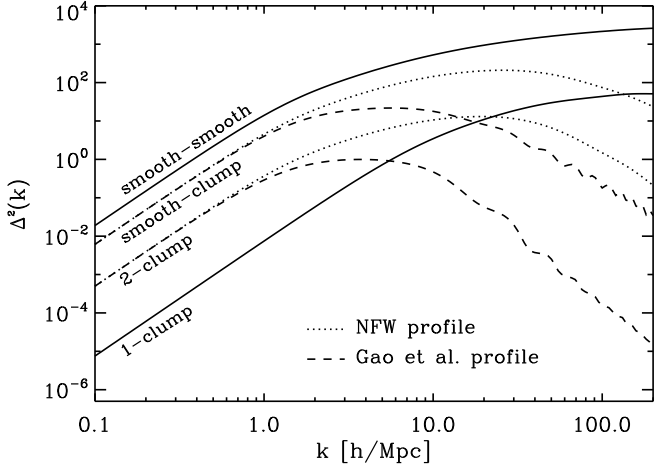, height=2.7in}
}
\caption{ Comparison of individual subhalo terms in the halo model.
The smooth-clump $\Delta_{sc}^2$ and 2-clump $\Delta_{2c}^2$ terms
depend on the distribution of subhalo centers $U_c$ within a host
halo, having a much lower amplitude at $k > 1 h $ Mpc$^{-1}$ when
$U_c$ has the cored isothermal profile of Gao et al. (2004) compared
with the cuspy NFW profile.  The smooth-smooth $\Delta_{ss}^2$ and
1-clump $\Delta_{1c}^2$ terms are independent of $U_c$.  
See text for parameters used in the model.
}
\label{fig:terms}
\end{figure}

Fig.~\ref{fig:terms} illustrates the contributions from the individual
terms in the halo model.  We use the NFW profile (truncated at the
virial radius) for the input host halo $U(k,M)$ and subhalo $u(k,m)$,
and the concentration $c(M)=c_0 (M/10^{14}M_\odot)^{-0.1}$ with
$c_0=11$ that we find to approximate the ART host halos and is
identical to Dolag et al (2004) except for a 15\% increase in
amplitude.  We use $c_0^{\rm sub}=3$ for the subhalos to take into
account tidal stripping but also compare different values in Fig.~4
below.  For the distribution of subhalo centers, $U_c(k)$, we compare
the profile of NFW with that of Gao et al. (2004), who find the number
of subhalos within a host halo's virial radius $r_v$ to be
\begin{equation}
   \frac {N(<x)}{N_{tot}} = {(1+ac)x^\beta \over 1+acx^\alpha}\,, 
   \quad x=r/r_v 
\end{equation}
where $a=0.244, \alpha=2, \beta=2.75, c=r_v/r_s,$ and $N_{tot}$ is the
total number of subhalos in the host.  Since this distribution at
small $r$ is shallower ($\propto r^{-0.25}$) than the inner part of
the NFW profile ($\propto r^{-1}$), its Fourier transform $U_c(k)$ at
high $k$ is about a factor of $10$ lower than that of the NFW profile.
This decrement results in a much lower $\Delta_{sc}$ and $\Delta_{2c}$
as shown in Fig.~\ref{fig:terms} (dashed vs dotted curves).  We find
the subhalo centers in the ART simulation to follow approximately the
distribution of Gao et al.~although there is a large scatter.

We use the mass function of Sheth \& Tormen (1999) for the host halos
$N(M)$ and a power law $n(m,M) \propto m^{-1.9}$ that well
approximates the subhalo mass function in the ART simulation.  The
latter is normalized so that the total mass of subhalos in a host halo
adds up to $f$ times the host mass $M$ ($f=0.14$ in Fig.~3).  To
compare the halo model with simulations, we set the lower limit on the
$P_{ss}$ integral to $10^{10} h^{-1}\ M_\odot$, which is the smallest
halo present in the simulation (about ten times the simulation
particle mass).  The lower limit on the outer integrals of $P_{sc}$,
$P_{1c}$, and $P_{2c}$ corresponds to the smallest halo that contains
substructure, which we set to the smallest halo that we considered for
erasing substructure in \S~2: $2 \times 10^{12} h^{-1}\ M_\odot$.
Similarly, the lower limit on the inner integrals of these terms is
set to the smallest subhalo that can be resolved (i.e. $10^{10}
h^{-1}\ M_\odot$).

\begin{figure}[t]
\centerline{ \epsfig{file=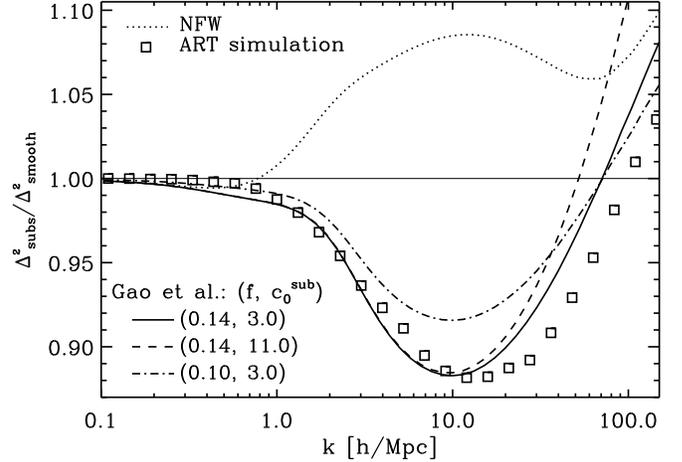, height=2.7in} }
\caption{ Same ratio of $\Delta^2$ as in Fig.~\ref{fig:pk} but
comparing simulation (symbols; same $10^{12.5} h^{-1} M_\odot$
curve in Fig.~\ref{fig:pk}) with halo model predictions (plain
curves).  The two agree qualitatively when the shallower distribution
of Gao et al. for subhalo centers $U_c(k)$ is used (bottom 3 curves)
in the halo model (but not for an NFW $U_c(k)$; dotted).  The
detailed model prediction depends on halo parameters: the solid curve
uses the same parameters as in Fig.~3; the dashed shows how a
larger subhalo concentration ($c^{\rm sub}= c_0^{\rm sub}(M/10^{14}
M_\odot)^{-0.1}$; $c_0^{\rm sub}=11$ vs 3) steepens the curve at high
$k$; the dash-dotted shows how a smaller subhalo mass fraction
$f$ (0.1 vs. 0.14) raises the dip.  }
\label{fig:halomod}
\end{figure}

Fig.~4 compares the sum of all the terms in the halo model with the
simulation result from Fig.~1.  As in Fig.~\ref{fig:pk}, we illustrate
the effects due to substructures by dividing out the power spectrum
from the original (smooth) halo model, i.e., $\Delta^2_{smooth}= k^3
P_{ss}/[(1-f)^2 (2\pi^2)]$, where $P_{ss}$ is given before in
eq.~(\ref{haloterms}).  Fig.~\ref{fig:halomod} shows that the halo
model is able to reproduce qualitatively the simulation results when
the subhalo centers in the halo model are assigned the shallower
distribution of Gao et al.  The feature of
$\Delta^2_{sub}/\Delta^2_{smooth} < 1$ at $1 \la k \la 100\,h$
Mpc$^{-1}$ is mainly caused by the drop in the smooth-clump term
relative to the smooth-smooth term at $k \ga 1\,h$ Mpc$^{-1}$ shown in
Fig.~\ref{fig:terms}.  The ratio $\Delta^2_{sub}/\Delta^2_{smooth}$
becomes $>1$ only at $k \ga 100\,h$ Mpc$^{-1}$ when the 1-clump term
finally takes over.  Fitting the halo model to actual simulation
results is clearly not exact in part due to the large scatters in the
properties of simulated halos, e.g., the concentration (for both hosts
and subhalos), the subhalo mass fraction $f$, and the maximum subhalo
mass in each host halo.  The halo model allows us to study the
dependence of clustering statistics on these parameters (see Fig.~4).
In addition, a number of effects are neglected in the current halo
model, e.g., tidal effects are likely to reduce the number of subhalos
(modeled by $U_c$ of Gao et al. here) as well as their outer radii
(not modeled here) towards host halo centers; a larger amount of
stripped subhalo mass may also be deposited to the inner parts of the
hosts, resulting in a radius-dependent subhalo mass fraction $f$
within the host.

Fig.~4 also shows that the halo model predicts the opposite effect due
to substructure (i.e.  $\Delta^2_{sub}/\Delta^2_{smooth} > 1$ at all
$k$) if the subhalo centers $U_c(k)$ are assumed to follow the NFW
distribution like the underlying dark matter.  The sign of this effect
is consistent with the previous subhalo model study of Dolney et
al. (2004), which assumed the same NFW profile for the subhalo centers
and the hosts and obtained a matter power spectrum that had a {\it
higher} amplitude for all $k$ when the substructure terms were
included.  Their results differ slightly from ours because of
different integration limits.  Subhalos in recent simulations like
that of Gao et al., however, show a much shallower radial distribution
in the central regions of the host halos, and inclusion of gas
dynamics appears to have little effect on the survivability of
subhalos (Nagai \& Kravtsov 2005).  The shallow distribution is
apparently due to tidal disruptions, even though the precise shape of
the distribution is still a matter of debate (e.g., Zentner et
al. 2005).  We have also experimented with a third distribution
$U_c(k)$ that has the NFW form but is less concentrated.  We are able
to bring $\Delta^2_{sub}/\Delta^2_{smooth}$ below unity only when the
concentration is reduced by a factor of more than 2.5, and only when
this reduction factor is increased to $\sim 100$ would we get a
comparable dip as the curves for ART simulation and Gao et al. in
Fig.~\ref{fig:halomod}.  It is interesting to see if we can mimic the
behavior of the ART simulation {\em without} using subhalos in the
halo model.  We try replacing the one-halo term, $P_{1h}$, by one that
is a simple superposition of a Gao et al.~profile and the usual NFW
profile.  This accounts for the fact that $\sim 90\%$ of the mass is
in a smooth NFW profile and that $\sim 10\%$ is in subhalos, which
follow a flatter profile.  One would not expect the high-$k$ regime to
agree as it is dominated by the subhalos (the 1-clump term,
specifically).  The intermediate range, $1 \la k \la 100\,h$
Mpc$^{-1}$, is dominated by the host halo itself, but we
find no similarity in this range either.  Subhalos are therefore needed
if the halo model is to recreate the ART results.

\section{Discussion}

The purpose of this work is to provide a physical understanding of the
effects of substructures on clustering statistics.  By experimenting
with dark matter substructures in a cosmological simulation with
$512^3$ particles, we have shown that the power spectra of matter
fluctuations and weak lensing shear can change by up to $\sim 12$\%
(and up to $\sim 24$\% in the bispectrum) if a significant amount of
substructures is not resolved in a simulation. When a larger mass
fraction of the host halos is in the form of lumpy subhalos, we find
the effect is to lower the amplitude of the matter and weak lensing
power spectra at the observationally relevant ranges of $k \sim 1$ to
$100\,h$ Mpc$^{-1}$ and $l \la 10^5$, and to raise the amplitude on
smaller scales.  A similar drop in power is also seen in our analytic
halo-subhalo model when the subhalo centers $U_c$ within a host halo
are distributed with a shallower radial profile than the underlying
dark matter (as expected due to tidal effects).  A way to understand
the drop involves looking at where the dense regions are.  When $U_c$
has an NFW form the subhalos basically trace the smooth background.
Thus, there is never a decrease in power when the smooth-smooth and
smooth-clump terms are added because dense regions are in nearly the
same relative positions.  When we use a shallower profile for $U_c$,
the subhalos are not as numerous in the denser inner regions of the
background halo.  This decrease in the overlap between dense clump
regions and the dense inner regions causes the drop in power.

We have quantified the effects of substructures on clustering
statistics by erasing substructures in an $N=512^3$ simulation.  An
important related question is whether $N=512^3$, single-mass
resolution simulations such as the one used in our study has
sufficient resolution to measure the power spectrum to the few-percent
accuracy required by future surveys.  Note that at least hundreds of
particles and force resolution of $\sim$ kiloparsec are required to
ensure subhalo survival against tidal forces, placing stringent
requirements on the dynamic range of simulations.  Multi-mass
resolution simulations designed for subhalo studies, on the other
hand, do not give reliable predictions for $P(k)$ on quasi-linear
scales due to compromised resolution outside highly clustered regions.
The fact that the curves in Figs.~1 and 2 continue to change at the
few-percent level each time the mass threshold is lowered by 0.5 dex
from $10^{12.5}$ to $10^{10.5}\,h^{-1} M_\odot$ suggests that subhalos
of $M\la 10^{10.5}\,h^{-1} M_\odot$ may still be affecting the power
spectra at a comparable level and that $N > 512^3$ would be required.
We also find $> 3$\% changes in $P(k)$ at $k\sim 10 h$ Mpc$^{-1}$ in
the halo model as the minimum subhalo mass in the integration limit of
eq.~(3) is lowered to $10^7 h^{-1} M_\odot$ (although the exact
predictions are sensitive to the slope of the subhalo mass function,
which is assumed to be $-1.9$ here.)  Careful convergence studies with
higher resolution aided by insight from this study and detailed
semi-analytic models for halo substructure will likely be needed to
determine $N$.

There are other challenges to predicting accurately the weak lensing
signal on single halo scales.  The effect of neutrino clustering could
cause a rise in weak lensing convergence of $\sim 1\%$ at $\ell\sim
2000$ (Abazajian et al.~2005).  Two recent groups have investigated
different aspects of baryon effects.  White (2004) found that baryonic
contraction and its subsequent impact on the dark matter distribution
is capable of causing an increase in the weak lensing convergence
power of a few percent at $\ell \ga 3000$.  Zhan \& Knox (2004), on
the other hand, use the fact that the hot intracluster medium does not
follow the dark matter precisely and predict an opposite effect: a
suppression of weak lensing power of a few percent at $\ell \ga 1000$.
Unlike the effects of substructure and neutrino clustering, the baryon
effects cause departures from the pure dark matter weak lensing signal
that only get larger with increasing $\ell$.

We thank M. Boylan-Kolchin, W. Hu, D. Huterer, C. Vale, and P. Schneider for useful
discussions.  BH is supported by an NSF Graduate Student Fellowship.
CPM is supported in part by NSF grant AST 0407351 and NASA grant
NAG5-12173.  AVK is supported by NSF grants AST-0206216 and 0239759,
NASA grant NAG5-13274, and the Kavli Institute for Cosmological
Physics at the University of Chicago.

\end{document}